\documentclass[preprint,aps]{revtex4}

\usepackage{graphicx}
\usepackage{subfigure}
\usepackage{amsmath}

\begin{document}

\title{Recovery of Galilean Invariance in Thermal Lattice Boltzmann Models for Arbitrary Prandtl Number}

\author{Hudong Chen, Pradeep Gopalakrishnan, Raoyang Zhang}
\address{Exa Corporation, 55 Network Drive, Burlington, MA 01803}

%\vspace{0.3in} 

\begin{abstract}
In this paper, we demonstrate a set of fundamental conditions required for the formulation of a thermohydrodynamic 
lattice Boltzmann model at an arbitrary Prandtl number.  A specific collision operator form is then proposed 
that is in compliance with these conditions. It admits two independent relaxation times, one for viscosity 
and another for thermal conductivity. But more importantly, the resulting thermohydrodynamic equations based on
such a collision operator form is theoretically shown to remove the well known non-Galilean invariant 
artifact at non-unity Prandtl numbers in previous thermal lattice Boltzmann models with multiple relaxation times.
\end{abstract}
\maketitle
%\vspace{0.3in} 

%\newpage

\section{Introduction}

Lattice Boltzmann Methods (LBM) has been matured as an advantageous method for computational fluid dynamics during past three 
decades\cite{Benzi,ChenDoolen}. A lattice Boltzmann system can be understood as a mathematical model involving a 
system of many particles, similar to that of the classical Boltzmann kinetic theory but involving 
only a discrete set of particle velocity values.  
There are two fundamental dynamical processes in a lattice Boltzmann model, namely, advection and collision. 
The collision allows particles to interact locally and to relax to a desired equilibrium distribution. 
The most popular model used for the collision process is the so called BGK operator involving only a single 
relaxation time parameter\cite{BGK,SChen,CCM,Qian}. Although the BGK has enjoyed many advantages and successes,
one of the apparent drawbacks is that all its transport coefficients, such as viscosity and thermal conductivity, are equal 
(subject to a constant scaling factor), resulting in unit Prandtl number.
There have been various attempts in the past to extend the collision process by introducing more than one
tunable relaxation parameter\cite{ChenMolvig,Multi,Than} in a generic linearized collision matrix form\cite{Benzi,Higu}. 
However, the outcome have been a mixed one. These extensions are able to allow a variable Prandtl number,
however they all suffer from generating a spurious non-Galilean invariant term (proportional to Mach number)
whenever Prandtl number is not unity.  Obviously such a spurious term is neither present 
in the physical thermohydrodynamics\cite{Huang}, 
nor in the lattice Boltzmann models with the BGK collision operator. 

In this paper we analytically describe the underlying origin of the problem that has plagued 
existing lattice Boltzmann models with multiple relaxation times\cite{Multi}. Furthermore, we present a new theoretical
procedure for formulating a collision operator that avoids such a problem. Consequently, the resulting 
thermohydrodynamic system with an an arbitrary Prandtl number will not contain aforementioned non-Galilean 
invariant artifact. The theoretical framework of the new formulation is based on projecting the collision 
operator onto an orthogonal moment basis of Hermite tensor polynomials, according to that presented in previous 
studies\cite{ShanYuanChen,Zhang,Multi}. However, unlike its predecessors in which the collision operator realizes 
moment fluxes in the absolute reference frame (i.e., a reference frame in which the lattice is at rest), 
the new formulation enforces conditions on moment fluxes in a relative reference frame with respect to the local fluid 
velocity\cite{HGrad1,HGrad2}. Most distinctly, the new formulation ensures the non-equilibrium
moment flux for thermal energy (temperature) in the relative reference frame as an eingen-vector of the collision
operator. Collision operators constructed in accordance 
to these conditions will not suffer from the aforementioned non-Galilean artifact. In particular, an explicit analytical
form of such a collision operator based on leading order Hermite polynomials
is proposed that admits two independently controllable parameters, 
one for viscosity and another for thermal diffusivity\cite{Multi}.  
We analytically show that the correct thermohydrodynamic equations are 
obtained for an arbitrary Prandtl number without the previously known spurious term. 
In addition, a numerical verification analysis is carried out for a simple benchmark case to demonstrate the consequence.  

\section{Some Basics on a Collision Process}        

A standard lattice Boltzmann equation (LBE) is conventionally expressed in the form below\cite{FHP1,Benzi,ChenDoolen},
\begin{equation}
%\label{BGK}
f_i({\bf x} + {\bf c}_i, t + 1) = f_i({\bf x},t) + \Omega_i({\bf x},t)
\label{eqn1}
\end{equation}
where $f_i({\bf x}, t)$ is the particle distribution function for velocity value ${\bf c}_i$ at $({\bf x},t)$.
Here the so called lattice units convention is adopted in which $\Delta t = \Delta x = 1$. 
The fundamental hydrodynamic (conserved) moments are constructed out of $f_i({\bf x}, t)$ by summing over all
the particle discrete velocity values $\{ {\bf c}_i , i = 0, \ldots, b\}$, 
\begin{eqnarray}
\sum_i f_i({\bf x},t) &=& \rho({\bf x},t)  
\nonumber \\
\sum_i {\bf c}_i f_i({\bf x},t) &=& \rho({\bf x},t) {\bf u}({\bf x},t)
\nonumber \\  
\sum_i e_i f_i({\bf x},t) &=& \frac{1}{2} \rho({\bf x},t) ({\bf u}^2({\bf x},t) + DT({\bf x},t))  
\label{moments}
\end{eqnarray}
where $e_i \equiv \frac{1}{2} {\bf c}^2_i$, and constant $D$ is the dimension of particle velocity space (depending on the
choice of a lattice type).
Macroscopic quantities $\rho({\bf x}, t)$, ${\bf u}({\bf x},t)$, and $T({\bf x}, t)$ denote mass density, fluid velocity 
and temperature at $({\bf x},t)$ in a fluid system, respectively. 
In eqn.(\ref{eqn1}), $\Omega_i$ represents a collision process among particles. 
For a lattice Boltzmann model with thermodynamic degree of freedom, conservation of energy 
is imposed along with that for mass and momentum, so that,
\begin{equation}
\sum_i \chi_i \Omega_i = 0
\label{consv}
\end{equation}
where $\chi_i = 1$, ${\bf c}_i$, or $e_i$.  (Note that we only considered particle kinetic energy for an ideal gas system).
In addition to obeying the conservation laws, a collision 
process drives a lattice Boltzmann system to a desired local equilibrium at some appropriate relaxation rate(s).
Symbolically,
\[ \Omega_i \; : f_i({\bf x},t) \rightarrow f^{eq}_i({\bf x},t) , \;\; i = 0, \ldots , b \]
where $f^{eq}_i({\bf x},t)$ denotes the local equilibrium distribution function, and it gives the same values as
$f_i({\bf x},t)$ for the three conserved hydrodynamic moments, eqn.(\ref{moments}).  
Most commonly the equilibrium distribution $f^{eq}_i({\bf x},t)$ is given as a polynomial function
of $\rho({\bf x},t)$, ${\bf u}({\bf x},t)$ and $T({\bf x},t)$\cite{Qian,ChenMolvig,ShanYuanChen,ChenShan}. 
Without affecting the discussions thereafter and for simplicity, we choose the following explicit form
\begin{eqnarray}
f^{eq}_i = \rho w_i [1 &+& \frac{{\bf c}_i \cdot {\bf u}}{T} + \frac{({\bf c}_i \cdot {\bf u})^2}{2T^2} - \frac{{\bf u}^2}{2T} 
\nonumber \\
&+& \frac{({\bf c}_i \cdot {\bf u})^3}{6T^3} - \frac{({\bf c}_i \cdot {\bf u}) {\bf u}^2}{2T^2} \nonumber \\
&+& \frac{({\bf c}_i \cdot {\bf u})^4}{24T^4} - \frac{({\bf c}_i \cdot {\bf u})^2 {\bf u}^2}{4T^3} + \frac{{\bf u}^4}{8T^2} + \cdots \; ] 
\label{feq}
\end{eqnarray}
which can be understood as an expansion of the asympototic exponential form\cite{ChenMolvig,ChenShan},
\begin{equation}
f^{eq}_i = \rho {\tilde w}_i \exp\left[- \frac{({\bf c}_i - {\bf u})^2}{2T}\right]
\end{equation}
where ${\tilde w}_i \equiv w_i \exp(e_i/T)$.
In the above, the weighting factor $w_i$ satisfies a set of generic conditions on moment tensors 
(valid for all lattice types and lattice Boltzmann models)\cite{ChenShan,ChenGoldOrszag},
\begin{eqnarray} 
\sum_i w_i \underbrace{{\bf c}_i \ldots {\bf c}_i}_{\text{n}} =  \begin{cases}
T^{\frac{n}{2}} \Delta^{n}, & n = 2, 4,...., 2N \label{eq.identity} \\
                                              0, & n = \text{odd integers}

\end{cases}
\label{iso}
%T^{\frac{n}{2}} \Delta^{n}, & n = 2, 4,...., 2N \label{eq.identity} \\
%                                                       0, & n = \text{odd integers} 
\end{eqnarray}
where $\Delta^{n}$ is the $n$-th order Kronecker delta function tensor (see \cite{ChenOrszag,ChenShan}).
For recovering correct thermohydrodynamics, $N\geq$4, 
and terms up to $O({\bf u}^4)$ in eqn.(\ref{feq})
need to be retained\cite{ShanYuanChen,NieShanChen}. 

For a fluid in the Newtonian flow regime in which deviations from local equilibrium are small, 
one may formally represent $\Omega_i$ in a linearized form\cite{Benzi},
\begin{equation}
\Omega_i = \sum_j M_{ij} (f_j({\bf x},t) - f^{eq}_j({\bf x},t))
\label{linear}
\end{equation}
where $M_{ij}$ is the collision matrix.
To define a fully closed mathematical description, an explicit form for $M_{ij}$ must be 
provided that satisfy some essential properties, such as conservation laws, symmetry, as well as relaxation 
to the equilibrium. The simplest of such an explicit form of the collision matrix is 
the so called ``BGK'' model\cite{BGK,SChen,CCM,Qian}
\begin{equation}
M_{ij} = - \frac{1}{\tau} \delta_{ij}
\label{bgk}
\end{equation}
where $\delta_{ij}$ is a standard Kronecker delta function.  Here the scalar $\tau$ is the 
only relaxation time responsible for all transport coefficients.  Obviously the resulting
Prandtl number $Pr$ ($ \equiv \mu c_p/k$) is unity.

One of the earliest attempts to construct an explicit collision operator for 
a variable Prandtl number was made nearly two decades ago\cite{ChenMolvig}, in that
a set of fundamental conditions was presented in terms of the following eigen-value relations,
\begin{eqnarray}
\sum_i {\bf c}_i {\bf c}_i M_{ij} &=& - \frac{1}{\tau_\mu} {\bf c}_j {\bf c}_j
\nonumber \\ 
\sum_i e_i {\bf c}_i M_{ij} &=& - \frac{1}{\tau_k} e_j {\bf c}_j
\label{eigen}
\end{eqnarray}
corresponding to the moments of fluid momentum and energy fluxes, respectively. 
Obviously these eigen-value relations are trivially satisfied by the BGK collision model
and $\tau_\mu = \tau_k = \tau$. 
It is a straightforward algebra to show that the resulting viscosity and the thermal conductivity 
are directly related to these two eigen values,
\begin{equation}
\mu = \left(\tau_\mu - \frac{1} {2} \right)\rho T \;,\;\; k = \left(\tau_k - \frac{1} {2} \right) c_p \rho T
\label{transp}
\end{equation}
where $c_p \equiv (D + 2)/2$. The ratio of viscosity and thermal conductivity defines the so called Prandtl number,
\[
Pr \equiv c_p \mu / k = \left(\tau_\mu - \frac{1}{2}\right)/\left(\tau_k - \frac{1}{2}\right)
\]  
More recently the eigen-value based concept was made more theoretically systematic through expansions in
moment space spanned by the Hermite polynomials\cite{Multi,Zhang}, briefly described below 
\begin{eqnarray}
\sum_j M_{ij}(f_j - f^{eq}_j) &=& -w_i \sum_{n=2}^N \frac{1}{\tau_n n!} {\cal H}^{(n)} (\boldsymbol{\xi}_i) {\bf a}^{(n)}
\nonumber \\ 
{\bf a}^{(n)}({\bf x},t) &=& \sum_j {\cal H}^{(n)}(\boldsymbol{\xi}_j)(f_j({\bf x}, t) - f^{eq}_j({\bf x}, t)) 
\label{CollH}
\end{eqnarray}
where $\boldsymbol{\xi}_i \equiv {\bf c}_i/\sqrt{T}$, and ${\cal H}^{(n)}$ is the standard $n$-th order 
(``discretized'') Hermite polynomial\cite{HGrad1,HGrad2,ShanHe,ShanYuanChen}.
Notice that the summation in eqn.(\ref{CollH}) starts from $n = 2$ because of the conservation laws.
A more explicit expression of eqn.(\ref{CollH}) for terms up to $N = 3$ is given below,
\begin{eqnarray}
&&\sum_j M_{ij}(f_j - f^{eq}_j) = -w_i [ \frac{1}{2\tau_\mu T} \left(\frac{{\bf c}_{i\alpha} {\bf c}_{i\beta}}{T} - {\bf \delta}_{\alpha\beta}\right) \odot {\bf \Pi}' 
\nonumber \\ 
&+& \frac{1}{6\tau_k T^2} \left(\frac{{\bf c}_{i\alpha} {\bf c}_{i\beta}{\bf c}_{i\gamma} }{T} - {\bf c}_{i\alpha}  {\bf \delta}_{\beta\gamma} - {\bf c}_{i\beta}  {\bf \delta}_{\alpha\gamma} - {\bf c}_{i\gamma}  {\bf \delta}_{\alpha\beta}\right) \odot {\bf Q}' ]   
\label{CollH2}
\end{eqnarray}
with
\begin{eqnarray}
{\bf \Pi}'({\bf x}, t) &\equiv& \sum_j {\bf c}_j {\bf c}_j (f_j({\bf x}, t)  - f^{eq}_j({\bf x}, t))
\nonumber \\
{\bf Q}'({\bf x}, t) &\equiv& \sum_j {\bf c}_j {\bf c}_j {\bf c}_j (f_j({\bf x}, t) - f^{eq}_j({\bf x}, t))
 \label{Fluxes}
\end{eqnarray}
being the non-equilibrium momentum and energy flux tensors, respectively. The sign ``$\odot$'' in eqn.(\ref{CollH2}) is a short notation
for the contraction operation between two tensors into a scalar. Note in eqn.(\ref{CollH2}),
we have renamed $\tau_2$ to $\tau_\mu$ and $\tau_3$ to $\tau_k$ because of their direct connections to
viscosity and thermal conductivity.
Due to orthogonality of the Hermite polynomials, collision matrix in eqn.(\ref{CollH2}) (and in eqn.(\ref{CollH}))
fully satisfies the eigen-value relations eqn.(\ref{eigen}), so that it directly leads to 
the resulting eqn.(\ref{transp}) for transport coefficients\cite{Multi}.  It is worth mentioning that 
the first part in eqn.(\ref{CollH2}), associated with the momentum flux tensor was discovered earlier\cite{Rot,Chopard}.

The theoretical analysis and the results presented above are clear and generally applicable for all conventional lattice Boltzmann models.
However, there has been one well known problem as pointed out in the previous section, the collision models with multiple relaxation times
formulated in this existing framework contain a spurious non-Galilean invariant term in the resulting energy 
equation whenever the Prandtl number is not unity.

\section{Recovering Galilean Invariance in A Collision Process}        

As we shall explain in this section, the root cause of the non-Galilean invariance problem is 
the representation of the collision operator based on the velocity moments in the absolute reference frame. From the continuum kinetic theory\cite{Huang}, we know that the fundamental hydrodynamic quantities are related to
the conserved moments defined in the relative reference frame with respect to the local fluid velocity ${\bf u}({\bf x},t)$.
For instance, the temperature is a local
root-mean-square of the particle relative velocity values,
\[ T({\bf x}, t) = \frac{2}{D} \langle ({\bf v} - {\bf u}({\bf x}, t))^2 \rangle, \; \;\; {\bf u}({\bf x}, t) = \langle {\bf v} \rangle \]
This in turn indicates that their corresponding moment fluxes must also be in accordance to the
relative reference frame representation\cite{HGrad1,HGrad2}. Therefore, the central task in formulating 
a ``hydrodynamically correct'' collision operator in the lattice Boltzmann models 
is to comply with such a fundamental requirement.

\subsection{Basic Properties in Relative Reference Frame}

The non-equilibrium momentum and energy tensor fluxes in the relative reference frame are defined as,
\begin{eqnarray}
{\tilde {\bf \Pi}}'({\bf x}, t) &\equiv& \sum_j {\tilde {\bf c}}_j {\tilde {\bf c}}_j (f_j({\bf x}, t)  - f^{eq}_j({\bf x}, t))
\nonumber \\
{\tilde {\bf Q}}'({\bf x}, t) &\equiv& \sum_j {\tilde {\bf c}}_j {\tilde {\bf c}}_j {\tilde {\bf c}}_j (f_j({\bf x}, t) - f^{eq}_j({\bf x}, t))
 \label{NonEQFluxes}
\end{eqnarray}
where ${\tilde {\bf c}}_j({\bf x}, t) \equiv {\bf c}_j - {\bf u}({\bf x}, t)$ 
is the relative particle velocity with respect to
the local fluid velocity ${\bf u}({\bf x}, t)$.
The energy flux vector in the relative reference frame is thus obtained from 
contracting the 3rd order tensor ${\tilde {\bf Q}}'({\bf x}, t)$:
\begin{eqnarray}
{\tilde {\bf q}}'_\alpha ({\bf x}, t) &\equiv& \frac{1}{2} {\tilde Q}'_{\alpha \beta\beta}({\bf x}, t) 
\nonumber \\
&=& \sum_j {\tilde e}_j {\tilde {\bf c}}_{j,\alpha} (f_j({\bf x}, t) - f^{eq}_j({\bf x}, t))
\label{SEflux}
\end{eqnarray}
where ${\tilde e}_j \equiv \frac{1}{2} {\tilde {\bf c}}^2_j$. 
In the above equation, Greek indices denote tensor components in Cartesian coordinates, 
and the summation convention is adopted for repeated Greek indices.

Due to the conservation laws, we easily see that the two non-equilibrium momentum fluxes (eqn.(\ref{Fluxes}) and eqn.(\ref{NonEQFluxes})) 
from the absolute and the relative reference frames are identical. In order words, the non-equilibrium momentum flux
is invariant no matter it is in the absolute or the relative reference frame.
\[ {\tilde {\bf \Pi}}'({\bf x}, t) = {\bf \Pi}'({\bf x}, t) \]
In contrast, however, the non-equilibrium energy flux vectors in the two reference frames are different, and are related according to 
\begin{equation}
{\tilde {\bf q}}'({\bf x}, t) = {\bf q}'({\bf x}, t) - {\bf u}({\bf x}, t) \cdot {\bf \Pi}'({\bf x}, t)
\label{qrelation}
\end{equation} 
where
\[ {\bf q}'({\bf x}, t) = \sum_j e_j {\bf c}_j (f_j({\bf x}, t) - f^{eq}_j({\bf x}, t)) \]
is the familiar non-equilibrium energy flux vector in the absolute reference frame.  This non-invariance
in non-equilibrium energy flux between the absolute and relative frames 
is the leading cause for the non-Galilean invariance artifact in
resulting thermohydrodynamics for a non-unity Prandtl number.

For recovering the right thermohydrodynamics from a lattice Boltzmann model with a variable Prandtl number, 
the key step is to construct a collision operator which ensures the conservation laws
as well as thermohydrodynamic fluxes in the relative reference frame. A set of fundamental
conditions required for such a collision operator is given below,
\begin{equation}
\sum_i {\tilde \chi}_i \Omega_i = 0
\label{consvR}
\end{equation}
where ${\tilde \chi}_i = 1$, ${\tilde {\bf c}}_i$, or ${\tilde e}_i$, and
\begin{eqnarray} 
\sum_i {\tilde {\bf c}}_i {\tilde {\bf c}}_i \Omega_i &=& - \frac{1}{\tau_\mu} {\tilde {\bf \Pi}}'
= - \frac{1}{\tau_\mu} {\bf \Pi}'
\nonumber \\  
\sum_i {\tilde e}_i {\tilde {\bf c}}_i \Omega_i &=& - \frac{1}{\tau_k} {\tilde {\bf q}}' 
= - \frac{1}{\tau_k} [{\bf q}' - {\bf u} \cdot {\bf \Pi}' ]
\label{Fcolli}
\end{eqnarray}
Clearly, eqn.(\ref{consvR}) is automatically satisfied by a lattice Boltzmann model (eqn.(\ref{consv})).
On the contrary, eqn.(\ref{Fcolli}) is not at all guaranteed from its counter part in the absolute reference frame.
In fact, one can verify that a collision operator admitting the original absolute 
reference frame based eigen-value relations, eqn.(\ref{eigen}),  
leads to a different result for energy flux below,
\begin{equation}
\sum_i {\tilde e}_i {\tilde {\bf c}}_i \Omega_i = - \frac{1}{\tau_k} {\bf q}' 
+ \frac{1}{\tau_\mu}{\bf u} \cdot {\bf \Pi}' 
\label{wrong} 
\end{equation}
Recall eqn.(\ref{qrelation}), we see that the eqn.(\ref{wrong}) is not equal to eqn.(\ref{Fcolli}) unless $\tau_\mu = \tau_k$.  
Also we can realize that the BGK collision operator trivially 
satisfies all these moment conditions in both reference frames.

Fundamental conditions, eqn.(\ref{Fcolli}), can also be casted in the form of the collision matrix $M_{ij}$, for momentum and energy fluxes in the relative reference frame into the following eigen-value relations (as compared to eqn.(\ref{eigen})) 
\begin{eqnarray}
\sum_i {\tilde {\bf c}}_i {\tilde {\bf c}}_i M_{ij} &=& - \frac{1}{\tau_\mu} {\tilde {\bf c}}_j {\tilde {\bf c}}_j
\nonumber \\ 
\sum_i {\tilde e}_i {\tilde {\bf c}}_i M_{ij} &=& - \frac{1}{\tau_k} {\tilde e}_j {\tilde {\bf c}}_j
\label{eigenR}
\end{eqnarray}
formally understood to be valid when projected onto the space spanned by $\{ f_j - f^{eq}_j \}$. 

In compliance with the relations in eqns.(\ref{consvR}-\ref{Fcolli}), we construct a collision operator, 
for example, with the following explicit form: 
\begin{eqnarray}
&&\sum_j M_{ij}(f_j - f^{eq}_j) = -w_i [ \frac{1}{2\tau_\mu T} \left(\frac{{\bf c}_{i\alpha} {\bf c}_{i\beta}}{T} - {\bf \delta}_{\alpha\beta}\right) \odot {\bf \Pi}' 
\nonumber \\ 
&+& \frac{1}{6\tau_k T^2} \left(\frac{{\bf c}_{i\alpha} {\bf c}_{i\beta}{\bf c}_{i\gamma} }{T} - {\bf c}_{i\alpha}  {\bf \delta}_{\beta\gamma} - {\bf c}_{i\beta}  {\bf \delta}_{\alpha\gamma} - {\bf c}_{i\gamma}  {\bf \delta}_{\alpha\beta}\right) \odot {\bf W}' ]   
\label{CollH2R}
\end{eqnarray}
Notice that the above collision form is identical to the previous form, eqn.(\ref{CollH2}),  except that the 3rd order moment tensor ${\bf Q}'({\bf x}, t)$ is replaced by ${\bf W}'({\bf x}, t)$.
The symmetric 3rd order moment tensor ${\bf W}'({\bf x}, t)$ is defined as,
\begin{equation}
{\bf W}'_{\alpha\beta\gamma} \equiv {\bf Q}'_{\alpha\beta\gamma} 
+ h \; [ {\bf u}_\alpha {\bf \Pi}'_{\beta\gamma} + {\bf u}_\beta {\bf \Pi}'_{\gamma\alpha} + {\bf u}_\gamma {\bf \Pi}'_{\alpha\beta} ] 
\label{Wtensor}
\end{equation}
where the scalar $h = 2\left(\frac {\tau_k}{\tau_\mu} - 1\right)$.
Using the fundamental lattice isotropy conditions eqn.(\ref{iso}), 
the proof for eqn.(\ref{CollH2R}) satisfying the conditions eqns.(\ref{consvR})-(\ref{Fcolli}) (and eqn.(\ref{eigenR})) 
is a straightforward algebra, and is not presented here.
One essential feature to be mentioned about the new collision operator is that, 
unlike the standard Hermite moment expansion in eqn.(\ref{CollH}) (and in eqn.(\ref{CollH2})), 
the exact parity between each Hermite basis function ${\cal H}^{(n)}$ 
and its coefficient ${\bf a}^{(n)}$ (moment of the same ${\cal H}^{(n)}$) is no longer held by eqn.(\ref{CollH2R}). 
The new coefficient is a function that also includes appropriate other (lower order) 
Hermite moments. Recovery of the correct thermohydrodynamics for arbitrary Prandtl numbers by the new collision 
operator form is theoretically shown in Appendix A.

\subsection{Numerical Verification}
The new theoretical formulation addresses thermal fluid flows in general. But in order to demonstrate the consequence, 
we have performed a numerical simulation of representative case in the linearized hydrodynamic regime using a 
two-dimensional 37 speed lattice model\cite{Multi}, corresponding to a ninth-order accurate Gauss-Hermite Quadrature. 
We have employed fourth order Hermite expansion of Maxwellian distribution for equilibrium\cite{ShanYuanChen,Multi} 
for accurate thermohydrodynamic representation. We perform a simple flow simulation with the following initial conditions, 
\begin{eqnarray}
u_x &=& 0 \nonumber \\
u_y &=& U_0 + \delta U_0 sin(nx) \nonumber \\  
T &=& T_0 + \delta T_0 sin(nx) \nonumber \\
\rho &=& \rho_0 \left [1 - (\delta T_0/T_0) sin(nx) \right] ;\quad n = 2\pi/L_x
\end{eqnarray}
A sinusoidal perturbation is given to the temperature on top of the lattice temperature $T_0=0.697953322019683$, 
which satisfies the isotropy condition given by eqn.(\ref{iso}). An inverse perturbation is given to the density to maintain 
the pressure ($\rho T$) approximately constant to avoid any acoustic modes. 
In order to exemplify the difference between the two collision models, we introduce a translational velocity $U_0$ 
(constant in space and time) to mimic a flow in a moving reference frame. A velocity shear is also added, 
which will result in non-zero momentum flux tensor $\Pi'$. The amplitude of the perturbations are kept at $\delta T_0=0.001$, 
and $\delta U_0=0.001$. The mean density $\rho_0$ is kept at one. All simulations are carried out in one dimensional grid, $L_x=100$, $L_y=1$,  
as there are no variations along the y-axis. Under such an initial condition the temperature diffusion of the 
perturbation follows a simple exponential decay,
\begin{equation}
T(t) = T_0 + \delta T_0 e^{-\kappa_0n^2t}; 
\end{equation}
where $\kappa_0 $ is the thermal diffusivity at mean temperature and density,
\begin{equation}
 \kappa_0  = \frac{k}{\rho_0 c_p} = \left(\tau_k - \frac{1} {2} \right) T_0 \nonumber
\end{equation}
Obviously, such a decay rate for a Galilean invariant model should be independent of the translational velocity value $U_0$. 

The simulations are carried out for different Prandtl numbers and translational velocity values for both 
previous collision model\cite{Multi}, eqn.(\ref{CollH2}), and current collision model eqn.(\ref{CollH2R}). 
For convenience, a small fixed viscosity value, $\mu = 0.00697953322019683$, corresponding to $\tau_\mu = 0.51$ is used for all the simulations. 
Different thermal diffusivity values are chosen to get the Prandtl number values ranging from 0.5 to 2.0. 
Figure 1 shows the ratio of the resulting thermal diffusivity from the simulation to the theoretical diffusivity $\kappa_0$, 
for different translational velocity values.  
One can see that the previous collision model exhibits an apparent non-Galilean invariant behavior 
in thermal diffusivity at non-unity Prandtl numbers, while the new model largely
removes this error. We have verified that the remaining error is in part related to a lack of isotropy in
the lattice velocity set higher than ninth order quadrature.

\begin{figure}[htbp]
\centering
\subfigure[]{
\includegraphics[width=\textwidth]{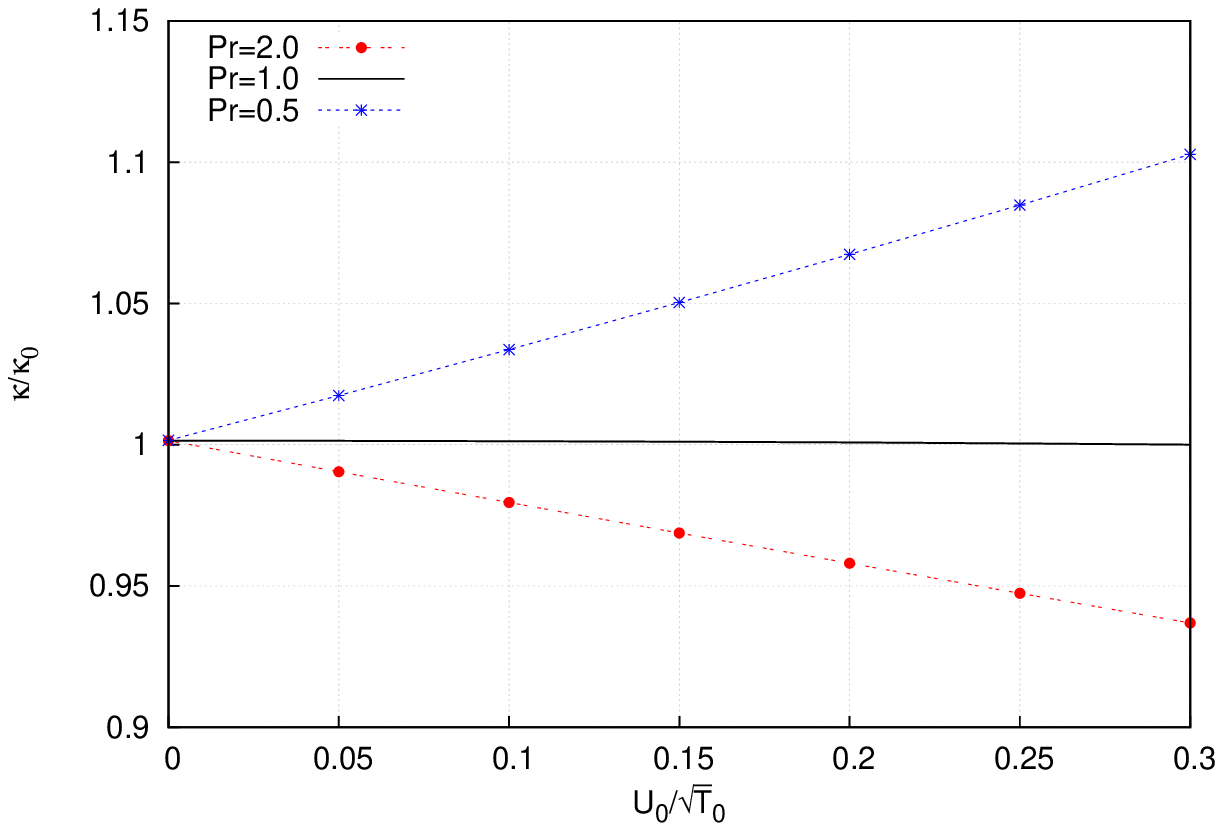}
}
\quad \quad
\subfigure[]{
\includegraphics[width=\textwidth]{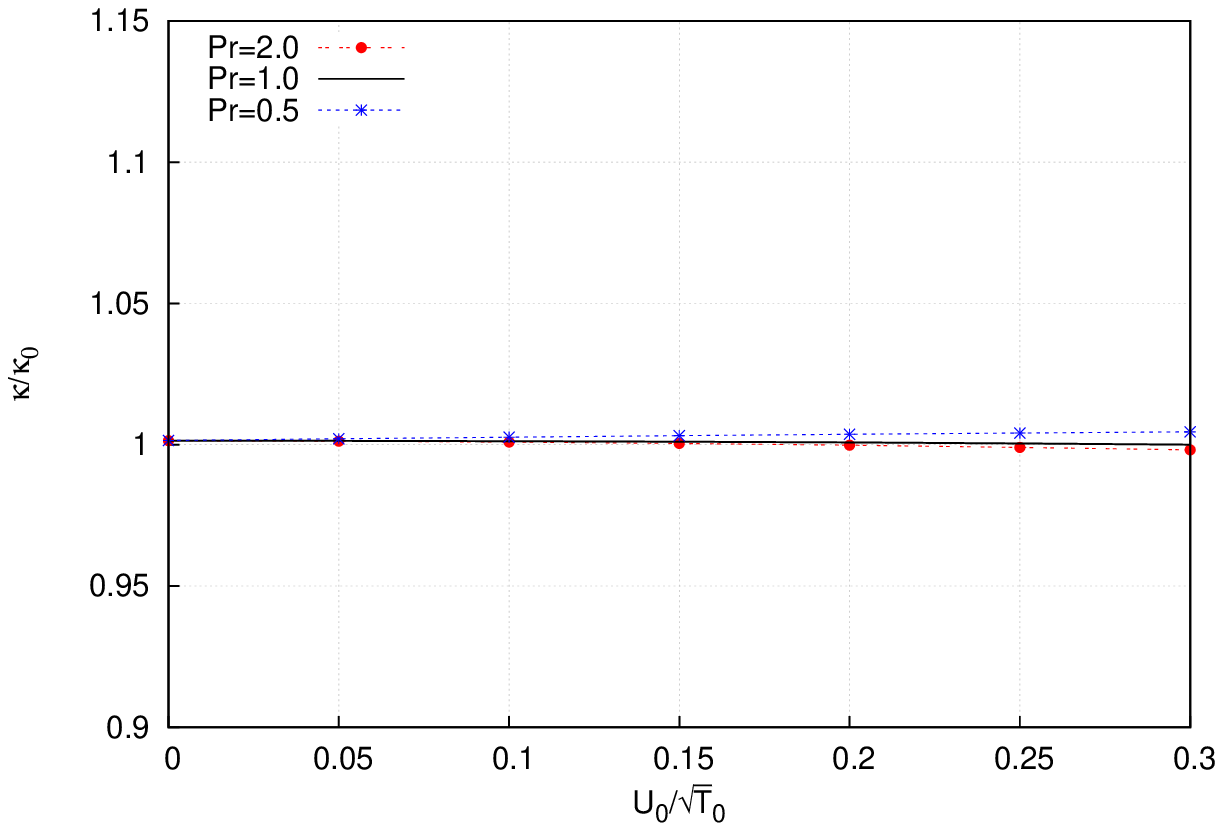}
}
\caption{Variation of thermal diffusivity $\kappa$ with translational velocity of a moving reference frame for
a) the previous collision operator and b)the new collision operator, where  
$\kappa_0$ is the theoretical diffusivity at mean temperature and density.}\label{fig1}
\end{figure}

\section{Discussion}

In this paper, we present a theoretical framework for formulating lattice Boltzmann models
resulting in the correct thermohydrodynamic equations with an arbitrary Prandtl number.  Unlike the previous 
approaches, lattice Boltzmann models derived under the new framework do not suffer from the well-known non-Galilean 
invariant artifact when Prandtl number is not unity. The critical difference of the present approach from previous 
ones is that the collision operator complies with a set of fundamental conditions pertaining to thermohydrodynamic 
fluxes (eqn.(\ref{Fcolli})) in the relative reference frame. 
A new explicit form of the collision operator is expressed that is
generally applicable for a class of lattice Boltzmann velocity types.  The latter are those that sufficiently meet 
the generic set of moment isotropy conditions, eqn.(\ref{iso}). It is important to point out that 
the non-Galilean invariance problem is not only a symptom of the specific Hermite expansion based formulations 
(e.g., eqn.(\ref{CollH}) or eqn.(\ref{CollH2})). Except BGK, this problem has been prevalent among most 
lattice Boltzmann models having multiple relaxation times in their collision process. The use of moments in the relative reference frame 
was previously proposed for formulating certain isothermal ``MRT'' type of LBM models\cite{cas1,cas2}. 
Recovery of correct thermohydrodynamics at arbitrary Prandtl numbers has not been met in previous LBM models. 

A few fronts in thermohydrodynamics may be expanded to beyond the current framework of collision operator representation. 
First of all, some potential energy or internal degrees of freedom may be incorporated so that the specific heat values 
are not restricted to the dimensions of lattices used\cite{ChenDoolen,Nie}. Care must be 
taken, however, in order to ensure the Galilean invariance via a proper Hermite expansion remains to be preserved (see
a related discussion in the paragraph below).  
Another extension is pertaining to boundary conditions, to ensure consistency with a thermohydrodynamic system with non-unity Prandtl numbers.  
Indeed, a (soild) boundary should also bear a Prandtl number effect in its thermodynamics (equilibrium and non-equilibrium) as well as 
the boundary momentum and heat fluxes. On the other hand, unlike collisions in a fluid flow domain, 
the issue of a relative reference frame may not arise if the boundary is stationary (i.e., ${\bf u} = 0$ at wall), and it may also
seem straightforward conceptually (albeit pontentially algebraically involved) for a moving boundary with a prescribed velocity. 
Another important point that deserves further investigations in the future: The current framework only addresses the Prandtl number 
related Galilean invariance issue associated with the pre-existing inadequate formulation of collision operators\cite{Multi}. 
With the new collision operator form, the correct full thermohydrodynamic equations are 
analytically shown to be recovered as long as a sufficient order isotropy (quadrature) (of ninth order or higher) 
is applied (see Appendix). Unfortunately, though significantly reduced,
the Galilean invariance artifact is still conspicuously observed numerically at finite translational velocity values in the resulting 
thermal diffusivity at non-unity Prandtl numbers, via the thermal decay test case in the previous section. Though simple,
it needs to be said that that the thermal decay problem is known to be far more non-trivial and delicate than that 
for a transverse shear fluid velocity\cite{Huang}. Nonetheless, such an observed small residual error ($\sim 0.1 \%$) in thermal diffusivity in our
numerical tests requires more understanding in the future.  One reason we know is that the higher order Hermite polynomials and lattice
isotropy beyound those for deriving the thermal diffusion term are exhibiting an higher impact at non-unity Prandtl numbers.  
This has been directly verified by either reducing or increasing quadrature orders in a lattice velocity set used in the test. 
On the other hand, one needs to find the theoretical origin of such an increased sensitivity at non-unity Prandtl numbers. 

Borrowing from the notion of gauge transformation, one may notice that
the explicit collision operator form eqn.(\ref{CollH2R}) presented in the preceding section is not unique for satisfying
the set of conditions eqns.(\ref{consvR})-(\ref{Fcolli}).  
For instance, more terms of higher order Hermite tensors may be added, similar to terms in
eqn.(\ref{CollH}) but not in eqn.(\ref{CollH2}).  Obviously, due to the orthogonality property of Hermite tensor polynomials, 
additional terms corresponding to higher order Hermite tensor polynomials will not alter the results for the lower order ones\cite{ShanYuanChen,Multi}.  
In particular, adding terms with Hermite tensor polynomials of order higher than 3 to the collision expression eqn.(\ref{CollH2R}) 
will not change its property in regard to the conditions eqns.(\ref{consvR})-(\ref{Fcolli}). Furthermore, 
such a systematic procedure of adding terms of higher order Hermite tensor polynomials
can also be used to extend the conditions of eqns.(\ref{consvR})-(\ref{Fcolli}) beyond the second and the third order moments. 
The key realization in formulating a collision operator via the Hermite expansion framework is to relax 
the parity constraints between each Hermite basis function and its coefficient,
so that the latter includes not only the moment of its own Hermite function but also 
appropriate moments from the lower order Hermite functions, as a result of
particle velocity in the relative reference frame. Collision operators satisfying an extended set of moment 
conditions may be relevant for fluid flows in broader regimes including complex and non-Newtonian fluid physics.
 
\appendix

\section{Deriving Thermohydrodynamic Equations}

To demonstrate that the new fundamental conditions eqns.(\ref{consvR}) - (\ref{Fcolli}) (or eqn.(\ref{eigenR}))
give rise to the correct thermohydrodynamic equations with a variable Prandtl number, we give a brief derivation below. 

Expanding eqn.(\ref{eqn1}) in Taylor series  in powers of spatial and time derivatives 
and neglecting terms beyond the 2nd order, we have
\begin{equation}
[\partial_t + {\bf c}_i\cdot \nabla + \frac{1}{2} (\partial_t + {\bf c}_i \cdot \nabla )^2] f_i({\bf x},t) \approx \Omega_i({\bf x},t)
\label{lpde}
\end{equation}
We then apply the standard multiple scale expansion procedure for time 
and spatial scales as well as for the distribution function\cite{CE,FHP1},
\begin{eqnarray}
\partial_t &=& \epsilon \partial_{t_0} + \epsilon^2 \partial_{t_1} \; ,  \;\; \; \nabla = \epsilon \nabla    
\nonumber \\
f_i &=& f^{eq}_i + \epsilon f^{(1)}_i + \epsilon^2 f^{(2)}_i + \cdots
\label{mexpan}
\end{eqnarray}
where $\epsilon$ denotes a small parameter.  The higher order parts of the distribution function ($f^{(n)} , \; n\geq 1$)
represent deviations from local equilibrium of various degrees, and having vanishing contributions to the conserved quantities,
\begin{equation}
\sum_i \chi_i f^{(n)}_i= 0 \; , \;\;\; n \geq 1
\label{consvNeq}
\end{equation}
with $\chi_i = 1$, ${\bf c}_i$, or $e_i$. Separating terms according to powers of $\epsilon$, so the
following set of hierarchical relations results
\begin{eqnarray}
& & \Omega_i(\{ f^{eq}\}, {\bf x}, t) = 0 \nonumber \\
& & (\partial_{t_0} + {\bf c}_i \cdot \nabla ) f^{eq}_i = \sum_j M_{ij} f^{(1)}_j \nonumber \\
& & \partial_{t_1} f^{eq}_i  + \frac{1}{2} (\partial_{t_0} + {\bf c}_i \cdot \nabla )^2 f^{eq}_i  + (\partial_{t_0} + {\bf c}_i \cdot \nabla ) f^{(1)}_i = \sum_j M_{ij} f^{(2)}_j
\label{hierac}
\end{eqnarray}
The first equation in eqn.(\ref{hierac}) above is trivially satisfied with a matrix operator representation.
We can then take summations of the conserved moments $\chi_i$. In particular, from the second equation 
in eqn.(\ref{hierac}) and applying the eigen-value relations, eqn.(\ref{eigenR}), we get
\begin{eqnarray}
\sum_i {\bf c}_i {\bf c}_i f^{(1)}_i = &-& \tau_\mu \sum_i {\bf c}_i {\bf c}_i (\partial_{t_0} + {\bf c}_i \cdot \nabla ) f^{eq}_i 
\nonumber \\
\sum_i e_i {\bf c}_i f^{(1)}_i = &-& \tau_k \sum_i e_i {\bf c}_i (\partial_{t_0} + {\bf c}_i \cdot \nabla ) f^{eq}_i 
\nonumber \\
&+& (\tau_k - \tau_\mu )\sum_i {\bf c}_i {\bf c}_i (\partial_{t_0} + {\bf c}_i \cdot \nabla ) f^{eq}_i 
\label{momt2}
\end{eqnarray}  
Notice the term proportional to $(\tau_k - \tau_\mu )$ in the second equation above would be missing if
the absolute reference frame based collision form, eqn.(\ref{CollH}), or the eigen-value relations eqn.(\ref{eigen}), were used instead.

After recombining terms of all orders, we get the overall set of equations for the conserved moments, 
\begin{eqnarray}
& & \partial_t \rho + \nabla \cdot (\rho {\bf u}) = 0
\nonumber \\
& & \partial_t (\rho {\bf u}) + \nabla \cdot [{\bf \Pi}^{(0)} + {\bf \Pi}^{(1)} ] = 0
\nonumber \\
& & \partial_t (\rho [\frac{D}{2} T + \frac{1}{2} {\bf u}^2]) + \nabla \cdot [ {\bf q}^{(0)} + {\bf q}^{(1)}] = 0
\label{contin}
\end{eqnarray}
where the momentum and energy fluxes of different orders are dependent solely on the equilibrium distribution function,
\begin{eqnarray}
{\bf \Pi}^{(0)} &\equiv& \sum_i {\bf c}_i {\bf c}_i f^{eq}_i
\nonumber \\
{\bf q}^{(0)} &\equiv& \sum_i e_i {\bf c}_i f^{eq}_i
\nonumber \\
{\bf \Pi}^{(1)} &\equiv& - (\tau_\mu - \frac{1}{2} ) \sum_i {\bf c}_i {\bf c}_i (\partial_{t_0} + {\bf c}_i \cdot \nabla ) f^{eq}_i 
\nonumber \\
{\bf q}^{(1)} &\equiv& - (\tau_k - \frac{1}{2} ) \sum_i e_i {\bf c}_i (\partial_{t_0} + {\bf c}_i \cdot \nabla ) f^{eq}_i 
\nonumber \\
& & + (1 - \frac{\tau_k - \frac{1}{2}}{\tau_\mu - \frac{1}{2}} ) {\bf u} \cdot {\bf \Pi}^{(1)} 
\label{fluxes12}
\end{eqnarray}
Once again, the term proportional to ${\bf u} \cdot {\bf \Pi}^{(1)}$ in the resulting ${\bf q}^{(1)}$ expression above 
would be missing if an absolute reference frame based formulation were used.  

The remaining derivations on $\sum_i {\bf c}_i {\bf c}_i (\partial_{t_0} + {\bf c}_i \cdot \nabla ) f^{eq}_i$ 
and $\sum_i e_i {\bf c}_i (\partial_{t_0} + {\bf c}_i \cdot \nabla ) f^{eq}_i$ 
are straightforward, since they no longer rely on properties of a collision operator and only 
depend on the form of the equilibrium distribution and its spatial and time derivatives (in the leading order).  
Using the 4th order form, eqn.(\ref{feq}), for $f^{eq}_i({\bf x}, t)$ and applying the generic 
lattice isotropy conditions, eqn.(\ref{iso}), up to $N = 4$, we obtain
\begin{eqnarray}
\sum_i {\bf c}_i {\bf c}_i (\partial_{t_0} + {\bf c}_i \cdot \nabla ) f^{eq}_i &=&
\rho T [2 {\bf \Lambda} - \frac{2}{D} {\bf I} \nabla \cdot {\bf u} ]
\nonumber \\
\sum_i e_i {\bf c}_i (\partial_{t_0} + {\bf c}_i \cdot \nabla ) f^{eq}_i &=&
\frac{D + 2}{2} \rho T \nabla T + {\bf u}\cdot \rho T [2{\bf \Lambda} - \frac{2}{D} {\bf I} \nabla \cdot {\bf u} ]
\label{derivfeq}
\end{eqnarray}
where the 2nd rank tensor ${\bf \Lambda}$ in a component form is
$\Lambda_{\alpha\beta} \equiv \frac{1}{2} [\partial_\alpha u_\beta + \partial_\beta u_\alpha ]$ being the rate of strain tensor.
Combining all the results above, we arrive at the final forms for the hydrodynamic fluxes,
\begin{eqnarray}
{\bf \Pi}^{(0)} &=& \rho {\bf u} {\bf u} + \rho T {\bf I}
\nonumber \\
{\bf q}^{(0)} &=& \frac{D + 2}{2} \rho T {\bf u} + \frac{1}{2} \rho {\bf u}^2 {\bf u}
\nonumber \\
{\bf \Pi}^{(1)} &=& - \mu [2{\bf \Lambda} - \frac{2}{D} {\bf I} \nabla \cdot {\bf u} ]
\nonumber \\
{\bf q}^{(1)} &=& - k \nabla T + {\bf u} \cdot {\bf \Pi}^{(1)}
\label{hydrofluxes}
\end{eqnarray}
with $\mu \equiv (\tau_\mu - \frac{1}{2})\rho T$ and $k \equiv (\tau_k - \frac{1}{2})\frac{D + 2}{2}\rho T$ being 
the resulting viscosity and thermal conductivity, respectively.
Eqns.(\ref{hydrofluxes}) are the same as that of the thermohydrodynamics of a Newtonian physical fluid. 
It is worthwhile to verify that collision operators such as eqn.(\ref{CollH}) 
obeying eqn.(\ref{eigen}) would lead to an incorrect form for the non-equilibrium energy flux below,

\begin{equation}
{\bf q}^{(1)} = - k \nabla T + \frac{1}{Pr} {\bf u} \cdot {\bf \Pi}^{(1)}
\end{equation}
As a consequence, it would result in a wrong temperature equation with an additional spurious non-Galilean invariant
term proportional to $(1 - \frac{1}{Pr})$.

\end{document}